\PassOptionsToClass{hyphens}{url}
\documentclass[sigconf,nonacm,urlbreakonhyphens]{acmart}
\usepackage{orcidlink}

\usepackage[frozencache]{minted}

\usepackage{flushend}

\usepackage{rotwein}
\hypersetup{breaklinks=true}
\pgfplotsset{compat=1.17}

\newcommand{\stefan}[1]{}
\newcommand{\lukas}[1]{}

\newcommand*{\crs}{\ensuremath{\mathsf{crs}}}
\newcommand*{\NIZK}{\ensuremath{\Pi}}
\newcommand*{\NIZKSetup}{\ensuremath{\mathsf{Setup}}}
\newcommand*{\NIZKProof}{\ensuremath{\mathsf{Proof}}}
\newcommand*{\NIZKVerify}{\ensuremath{\mathsf{Verify}}}

\makeatletter
\let\@float@c@listing\@caption
\makeatother

\title{Extending Expressive Access Policies with Privacy Features}
\titlenote{This is the full version of a paper which appears in 21th IEEE International Conference on Trust, Security and Privacy in Computing and Communications (TrustCom 2022), December 9-11, 2022, Wuhan, China, IEEE. \url{http://www.ieee-hust-ncc.org/2022/TrustCom/accepted_paper.html}}

\author{Stefan More}
\orcid{0000-0001-7076-7563}
\email{stefan.more@iaik.tugraz.at}
\affiliation{%
  \institution{Graz University of Technology}
  \city{Graz}
  \country{Austria}
}

\author{Sebastian Ramacher}
\orcid{0000-0003-1957-3725}
\email{sebastian.ramacher@ait.ac.at}
\affiliation{%
  \institution{AIT Austrian Institute of Technology}
  \city{Vienna}
  \country{Austria}
}

\author{Lukas Alber}
\email{lukas.alber@iaik.tugraz.at}
\affiliation{%
  \institution{Graz University of Technology}
  \city{Graz}
  \country{Austria}
}

\author{Marco Herzl}
\email{herzl@student.tugraz.at}
\affiliation{%
  \institution{Graz University of Technology}
  \city{Graz}
  \country{Austria}
}

\begin{document}

\begin{abstract}
Authentication, authorization, and trust verification are central parts of an access control system.
The conditions for granting access in such a system are collected in access policies.
Since access conditions are often complex, dedicated languages -- policy languages -- for defining policies are in use.

However, current policy languages are unable to express such conditions having privacy of users in mind.
With privacy-preserving technologies, users are enabled to prove information to the access system without revealing it.

In this work, we present a generic design for supporting privacy-preserving technologies in policy languages.
Our design prevents unnecessary disclosure of sensitive information while still allowing the formulation of expressive rules for access control.
For that we make use of zero-knowledge proofs (NIZKs).
We demonstrate our design by applying it to the TPL policy language, while using SNARKs.
Also, we evaluate the resulting ZK-TPL language and its associated toolchain.
Our evaluation shows that for regular-sized credentials communication and verification overhead is negligible.
\end{abstract}

\maketitle

\section{Introduction}

Trust verification, authentication, and authorization are core concepts of access management.
Conditions out of all three concepts specify whether access can be granted.
They can involve the user, resources, the requested operation, and facts from the environment.
The conditions together form a set of rules called \emph{policy}.
While a policy may be simple and only define a single bit that grants access to a resource, often its conditions are complex and require an elaborate implementation of multiple intertwined checks into access control systems.
Furthermore, with many clients and services interacting, it can become quite tedious to implement different variants of access logic. %

\textbf{Access policy languages}~\cite{DBLP:conf/sp/BlazeFL96,DBLP:conf/itrust/Yao03,DBLP:conf/icws/YuanT05,DBLP:conf/secrypt/CoiO08,DBLP:conf/trustbus/AlmWP09,DBLP:conf/ifiptm/ModersheimSWMA19} enable the codification and re-use of access policies while decoupling them from the deployed access control systems.
Furthermore, policy languages offer a higher level of abstraction that facilitates the design of policies without requiring concrete insights into the implementation of the underlying access control system.
An example identity management model is Self-Sovereign Identity (SSI)~\cite{allen2016}.
Since this model often involves complex access policies, previous research has already addressed the integration of policy languages into the SSI model~\cite{DBLP:conf/trustcom/BelchiorPP0VG20,DBLP:conf/openidentity/AlberMMS21}, but neglects the topic of privacy.

To better highlight the privacy issues we aim to tackle, we first summarize the \emph{policy-based and SSI-based access control} model:
First, a policy is defined by a domain expert, codifying rules for authentication, authorization, and trust verification.
This policy is then stored at a \emph{service provider (SP)}.
As soon as a \emph{user} wants to access a resource or consume a service at this SP, they need to show that they fulfill the access policy.
For that the user relies on self-sovereign identity attributes, which they receive from an issuer or identity provider in the form of \emph{digital credentials}.
The user then stores their credentials in a \emph{digital identity wallet}~\cite{BlazLukasThomasIdentityWalletPaper}.
To authenticate, the user bundles the needed credentials to its service request and sends it to the SP.
After receiving the request, the SP uses a \emph{policy interpreter} to verify the incoming user request by applying the access policy.

However, this approach suffers from \textbf{privacy issues}.
Users are often in possession of credentials that certify numerous attributes.
When showing a credential to an SP, users reveal all attributes to the verifier, which is often neither desirable nor necessary to fulfill an authentication request.
By integrating privacy-preserving technologies into the access control process, users are enabled to only reveal a subset of the attributes or even prove a statement without revealing any attribute at all.
For example, by introducing Camenisch-Lysyanskaya (CL) signatures~\cite{DBLP:conf/scn/CamenischL02} into W3C's verifiable presentations~\cite{w3cVC}, support for privacy-preserving showings is achieved.
Those features are well-understood in the field of attributed-based credentials~\cite{DBLP:journals/cacm/Chaum81,DBLP:journals/cacm/Chaum85} and have been studied for SSI systems~\cite{DBLP:journals/csr/MuhleGGM18,DBLP:conf/icics/AbrahamHOR19,DBLP:conf/trustcom/AbrahamKMRS21}.

The proliferation of privacy-preserving  in combination with policy languages also enable new uses cases.
For example, in data marketplaces~\cite{DBLP:conf/primelife/KochKPR20,DBLP:conf/IEEEares/MoreA22} they allow data owners to define access policies based on the data sellers credential.
Thereby, privacy beyond the privacy of the shared data in such a marketplace may be achieved.

\subsection{Challenges}

While the above-mentioned research and standards are helpful, integrating privacy-preserving technologies into policy-based access control systems is not straightforward. Several challenges need to be addressed:
How should sensitive attributes be marked hidden in a policy language?
Which statements on the hidden attributes need to be revealed?
How is the user informed on the statements they need to prove?
How can a user-side wallet implementation support all the possible proof types different SPs may ask for?
Which privacy-preserving technologies can help to overcome these challenges while being flexible enough to preserve the expressiveness of the policy languages?

\subsection{Our contribution}

In this work, we close the gap by extending policy language systems with privacy features using zero-knowledge proofs.
Our contribution is as follows:

\textbf{Privacy-preserving Policy System:}
We introduce a generalized approach for extending existing policy language-based access control systems relying on SSI. %
For convenience, we will refer to the latter as \emph{policy systems}.
The policy's author codifies which statements the user should prove and which information needs to be revealed to receive access.
The author then uses our \emph{policy compiler} to derive a presentation request they provide to users.
The presentation request informs the user about the attributes they need to reveal and statements they need to prove.
That enables users to only provide the required attributes and statements, and hide the rest of the credential data.
While we focus our implementation and evaluation on the SSI model, the design itself is generic.
It can be applied to various policy systems and identity management models, enriching them with privacy features.

\textbf{Implementation:}
To show the feasibility of our design, we provide a concrete instantiation of our policy system extension.
First, we adapt the TPL trust policy system~\cite{DBLP:conf/ifiptm/ModersheimSWMA19} to enable privacy-preserving access control.
Second, we extend the syntax of TPL policies to allow denoting whether attributes should be revealed, as well as which statements should be proven without revealing additional information.
Third, we automatically compile the policies into the corresponding circuits for SNARK-based zero-knowledge proofs~\cite{DBLP:conf/innovations/BitanskyCCT12}.
Users later execute these circuits to generate proofs of attributes without revealing their value.

\textbf{Evaluation:}
Finally, we evaluate our implementation and discuss the overhead introduced by the additional computations.
We perform the evaluation for different commitments (SHA256, Poseidon) and two elliptic curves (ALT-BN128, BLS12-381).
We observe that the duration of the one-time setup and the computation of a proof depend on the choice of both commitment and curve.
However, only the verification is time-sensitive, since it needs to happen in real-time during the user's request for access.
We found that the verification introduces a negligible performance overhead. %
Since SNARKs are non-interactive, also the access verification is non-interactive.
Thus, a privacy-preserving showing requires no additional communication rounds between the user and the SP. %

\subsection{Related Work}
\label{sec:related-work}

Decentralized services like IPFS\footnote{\url{https://ipfs.io/}, accessed on 2022-07-02} store user data redundantly on one or more nodes. However, if the data is sensitive, it should not be accessible arbitrarily. Therefore, Prünster et al.~\cite{DBLP:conf/nss/PrunsterZP20} show an approach on how to outsource data protected by an access policy without needing to involve the data owner and thus
ensure decentralization. Roughly speaking, the sensitive data's encryption key is split into several parts stored on different, selected nodes. On access, these nodes evaluate the accessing party's attributes using a policy. If all agree on granting access, the encryption key can be recovered from the key shares, and the data turns accessible. All in all, their fully decentralized ABAC
implementation focuses on decentralized user data access (compared to service access) and uses the eXtensible Access Control Markup Language (XACML\footnote{\url{http://docs.oasis-open.org/xacml/3.0/xacml-3.0-core-spec-os-en.html}}) standard as policy language.

Belchior et al.~\cite{DBLP:conf/trustcom/BelchiorPP0VG20} propose a Self-Sovereign Identity based access control (SSIBAC) for service providers. It leverages conventional attribute-based access control using the attributes in SSI credentials, profiting from its decentralized nature. SSIBAC uses the XACML standard\footnote{\url{http://docs.oasis-open.org/xacml/3.0/xacml-3.0-core-spec-os-en.html}}
for policy specification. Their implementation is based on W3C Verifiable Credentials~\cite{w3cVC}, Hyperledger Indy\footnote{\url{https://github.com/hyperledger/indy-sdk}, accessed on 2022-07-03} and Aries\footnote{\url{https://github.com/hyperledger/aries}, accessed on 2022-07-03} for communication and distributed storage.
While they mention the possibility of introducing privacy-enhancing technologies, they do not discuss the integration into an access policy language. %

The ABC4Trust project focused on privacy-enhancing attribute-based credentials (ABC) and can be seen as preliminary work for the W3C VC standard~\cite{w3cVC} and modern SSI~\cite{DBLP:conf/trustbus/SabouriKR12,DBLP:books/daglib/p/BichselCDEKKLNPPRS15,DBLP:books/daglib/p/BichselCDEKLNP15}.
In their approach, the verifier sends a so-called presentation policy to the user.
This policy states the conditions the user has to fulfill in order to access the service.
On the user-side, the ABC engine then matches the needed attributes, and finally, a presentation token is created consisting of cryptographic evidence that the user satisfies the policy.
This proof can later be verified for access control purposes.
Since the whole policy must be proven through the presentation token, it can not contain any conditions that are difficult or impossible to translate to a cryptographic proof (e.g., online lookup for trust verification).
Additionally, the project supports a limited list of functions for use in predicates on private attributes.\footnote{\url{https://abc4trust.eu/download/Deliverable_D2.2.pdf}, Section 4.4.3}
Thus, policies in ABC4Trust enable some privacy features but are limited in their flexibility.
\stefan{clarify to what extend this is true}

\lukas{note: Programmable Credentials1996~\cite{DBLP:conf/sp/BlazeFL96}, Fidelis2003~\cite{DBLP:conf/itrust/Yao03}}
\lukas{Other (trust/access) policy stuff; cf. other TPL papers}

\section{Background}
\label{sec:background}

\subsection{Non-Interactive Zero-Knowledge Proof Systems}
\label{sec:snarks}

Non-interactive zero-knowledge (NIZK) proof systems represent powerful tools that enable a prover to convince a verifier of the validity of a statement without revealing any other information. For an NP-language \(L \subset X\) and a statement \(x \in X\), a prover can present a proof to the verifier that \(x \in L\), e.g., there exists a witness \(w\) such that \(x \in L\). No other information about the witness \(w\) is leaked to the verifier.
 Formally, let \(R\) be the associated witness relation such that \(L = \{ x \in X \:|\: \exists w\colon R(x, w) = 1\}\). A non-interactive proof system \(\NIZK\) consists of algorithms \(\NIZKSetup(1^\kappa)\) producing a common reference string \(\crs\), \(\NIZKProof(\crs, x, w)\) taking a statement \(x \in X\) and a witness \(w\), and outputting a proof \(\pi\), and \(\NIZKVerify(\crs, x, \pi)\) taking a statement \(x\) and a proof \(\pi\), and outputting the verification status.
We require such a proof system to be \emph{complete} (i.e., all proofs for statements in the language verify), \emph{sound} (i.e., a proof for a statement outside the language verifies only with negligible probability), and \emph{zero-knowledge} (i.e., a proof reveals no information on the witness).

Succinct Non-Interactive Arguments of Knowledge (\textbf{SNARKs}) are one of such systems and of particular interest to us, as they come with small proofs that are independent of the witness size and allow for fast verification. With the seminal work of Groth~\cite{DBLP:conf/eurocrypt/Groth16}, SNARKs have been improved in various directions. To reduce the trust assumptions necessary for the generation of the common reference string (CRS) \(crs\), subversion-resistant and updatable versions have been investigated~\cite{DBLP:conf/crypto/GrothKMMM18,DBLP:conf/ccs/AbdolmalekiRS20}.
SNARKs have been extended with stronger security notions such as strong simulation-sound extractability.
Toolsets including ZoKrates~\cite{DBLP:conf/ithings/EberhardtT18}, arkworks\footnote{\url{https://arkworks.rs}, accessed on 2022-07-07} or xJsnark~\cite{DBLP:conf/sp/KosbaPS18} provide compilers to turn arbitrary programs into circuits suitable for SNARKs or implement building blocks to help with the design of suitable circuits.

In this paper, we build our system on \textbf{ZoKrates}. It offers a high-level language syntax akin to Python with static typing. It allows implementing functions and programs that represent statements to be proven with a NIZK. Internally, the program will be represented as rank-1 constraint systems to be consumed by bellman\footnote{\url{https://github.com/zkcrypto/bellman}, accessed on 2022-07-01} which implements Groth's SNARK~\cite{DBLP:conf/eurocrypt/Groth16}.

\subsection{TPL Policy System}

\begin{figure}[htbp]
\centerline{\includegraphics[]{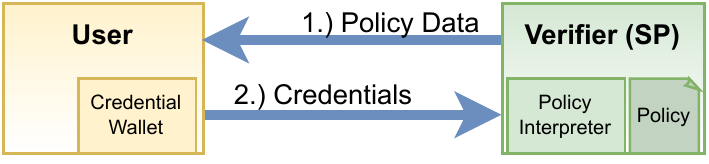}}
\caption{High-level architecture of an access process using a policy system.}\label{fig:highlevel}
\end{figure}

A policy language enables the SP to specify precise conditions based on which access to a resource can be granted.
We build upon the TPL language introduced by Mödersheim et al.~\cite{DBLP:conf/ifiptm/ModersheimSWMA19}. Together with its toolchain, we call it a policy system. This system enables an expressive definition of access policies. Subsequently, it allows for automated decisions on whether incoming access requests can be accepted. The expressiveness and the flexibility of the system allow it to be used in different a variety of SPs.
The system has its origin in the LIGHTest project~\cite{DBLP:conf/openidentity/BrueggerL16}
It supports DNS-based trust scheme verification~\cite{DBLP:conf/openidentity/Rossnagel17} to derive trust relations and establish trust in credentials. Later, Alber et al.~\cite{DBLP:conf/openidentity/AlberMMS21} added SSI concepts as an additional trust anchor option. The authoring of policies can happen natively in TPL, which has a  syntax similar to Prolog.
Additionally, non-technical domain experts are provided with an authoring tool with a graphical user interface~\cite{DBLP:conf/openidentity/ModersheimN19,DBLP:conf/openidentity/WeinhardtP19,DBLP:conf/icissp/WeinhardtO19}.

A TPL policy (see \Cref{tpl:basic-w3c} for an example) is a set of several rules in the form Horn clauses: $p(t) :- q_1(u_1),\dots,q_n(u_n)$.
A rule evaluates to true if all queries $q_i(u_i)$ return true.
A predicate $p$ can be defined by multiple rules (of the same name) and evaluates to true if any of the rules are satisfied.
To check whether a specific policy is met, the TPL interpreter evaluates a solution for query $p(s)$.
If it is fulfilled for the provided input data (e.g., identity data), the interpreter returns true.
The interpreter evaluates the query by searching for suitable rules for substitution.
If a query $p(s)$ and predicate $p(t)$ share an unifiable $s$ and $t$, a unifier is calculated and applied to the subqueries of the predicate's rule.
Subsequently, all subqueries are evaluated in the same way.
In case of a subquery returning false, another rule needs to be found for substitution. If all substitutable rules are exhausted, and no solution can be found, false is returned.

Subqueries are evaluated recursively until a ground truth is found.
Such truth can be a relational operation or a built-in predicate.
The TPL backend system handles all the built-in predicates. They help, for example, with server lookups for trust information discovery in eIDAS~\cite{DBLP:conf/openidentity/WagnerWMH19} and SSI~\cite{DBLP:conf/openidentity/AlberMMS21}.

The entry point of a policy is the \emph{accept} predicate.
To check if an incoming presentation is acceptable, the \emph{accept} predicate is initially called through a corresponding query.
The presentation token itself is the only input argument.

\begin{listing}[h]
\begin{minted}[fontsize=\footnotesize, frame=single, framesep=5pt]{prolog}
accept(Presentation) :-
 set_format(Presentation, w3cVP),
 extract(Presentation, verifiableCredential, Cred),
 set_format(Cred, w3cVC),
 extract(Cred, issuerDID, DIDissuer),
 checkQualified(DIDissuer),
 checkSig(Cred, DIDissuer),

 extract(Cred, credentialSubject, Subject),
 extract(Subject, date_of_birth, Birthdate),
 calculateAge(Birthdate, Age), Age >= 18,

 extract(Subject, username, Username),
 print(Username).
\end{minted}
\caption{Example TPL policy for W3C Verifiable Credentials, with the trust-check omitted for clarity.}
\label{tpl:basic-w3c}
\end{listing}

\section{Concept}
\label{sec:concept}

In this section, we describe the design of our access policy system with privacy-preserving features.
In \Cref{sec:implementation} we discuss a concrete instantiation of this design.

Before describing the different components, actors and the process, we will present the high-level idea of our concept.
\subsection{High-Level Idea}
\textbf{Preliminary: Commit-sign-proof Credentials}
One common approach (cf. \cite{DBLP:conf/scn/CamenischL02,DBLP:journals/joc/FuchsbauerHS19}) to design attribute credentials is to first commit to the attributes.
This commitment is then signed by the issuer (i.e., identity provider).\footnote{The signature and commitment may coincide, but for giving an intuition, we consider them distinct.}
For privacy-preserving showings, the user later proves consistency of any revealed attribute with respect to the commitment.
The latter proof is combined with a proof of knowledge of a signature on the commitment, or by directly providing the signature to the verifier.

\textbf{Compiling Access Policies into NIZK Proofs}
As in other systems with access control, %
rules that have to be satisfied must be represented as program logic, forming a policy.
Hence, we extend the concept of commit-sign-proof credentials with the possibility for a policy designer to codify such access rules.
For any committed-to attribute, we enable the policy designer to decide whether the user needs to reveal an attribute to the service provider or whether it is sufficient to convince the verifier that an attribute satisfies a policy rule without revealing it. %
Our system {automatically} informs the user about the policy and compiles it into the corresponding NIZK proofs.
That is, the user proves the consistency of revealed attributes with the public commitment.
Similarly, for all rules involving hidden attributes, the user also generates proofs of knowledge of these attributes and that they satisfy the specified rules.

In our system, we allow the policy to express rules with respect to any credential format:
attributes that are encoded in some form of data structure that has some public reference value.
The latter may consist of a signed commitment or a signature directly over the attributes.
The statement for the proof system is then built accordingly.

\subsection{Components}

Our system consists of the following actors/components:

\textbf{User (Prover):}
The actor that wants to access a resource or consume a service at the SP, and needs to authenticate to do so.
Their identity attributes are stored in form of \textbf{digital credentials} in a \textbf{digital identity wallet}~\cite{BlazLukasThomasIdentityWalletPaper}.
Part of the wallet, the user's system is also a \textbf{policy client}, which prepares a \textbf{presentation token} that satisfies a \textbf{presentation request}.

\textbf{Service Provider (Verifier):}
The SP is the actor (or their system) which provides access-protected services or resources to users.
To control who can access a resource and how users are authenticated.
The administrator of the SP creates a \textbf{policy} that encodes access control and trust rules.
The SP uses a \textbf{policy compiler} to generate the presentation request, which they provide to the user.
After receiving a service request, the SP uses a \textbf{(extended) policy interpreter} to verify the request.

\subsection{Phases}

We now discuss the entire process of our concept.
We split the process into the following phases:
\begin{enumerate*}
\item The setup of the cryptographic system, authoring of a policy, and publishing of the presentation request,
\item computing of a presentation token, and finally
\item the verification of the presentation token.
\end{enumerate*}

\textbf{(1) Setup System and Policy:}
If the employed proof system requires a specific setup,
performing it is the first step.
For example, when using SNARKs, a trusted third party generates a CRS.
The so-obtained common material is published and retrieved by all system participants.

During the setup phase, a \emph{policy} is authored by the service provider (SP).
This policy encodes the rules a user of the SP needs to fulfill to use the service or access a resource.
Depending on the nature of an SP, there can be different policies for different services or resources.

Authors of this policy are either technical personnel or domain experts of the SP.
Although an author without technical knowledge but with domain expertise can use a graphical policy authoring tool to create the policy~\cite{DBLP:conf/openidentity/WeinhardtP19,DBLP:conf/icissp/WeinhardtO19}, a policy is, in the end, always encoded in machine-readable form. %
As part of the policy, the author specifies which attributes a user has to provide and what credential types the SP accepts.
Additionally, the policy author defines two types of rules the user attributes have to satisfy, which are differentiated by whether they operate on private attributes or revealed attributes.
Our system later transforms the first set of rules into statements for the NIZK proof system.
Thus, the user can prove that their credentials fulfill these rules without revealing the values of the attributes.
The second set of rules operates on attributes the user must reveal to the SP.
These rules are used when the SP requires the attributes for further processing or trust management.
Depending on the NIZK proof system, the SP at this stage also compiles parts of the policy into an intermediate representation for the policy client.

The encoded policy, together with metadata about the service, forms the \emph{presentation request} for a specific service.
Before initiating a service request, users need to know what data they are required to provide.
Therefore, the presentation request is published by the SP.

\textbf{(2) Authentication at SP:}
When user want to access a service or consume a resource, they have to authenticate with the SP.
To do so, they first retrieve the corresponding presentation request from the SP's website or another form of a service catalog.

The user then extracts the policy from the presentation request and executes it using the policy client.
While doing so, the client retrieves the respective credentials from the user's identity wallet~\cite{prologABCmatcher}.
For the rules on public attributes, the client extracts the attributes from the credential, thereby revealing their values.
It then computes a NIZK statement that proves that the value was indeed extracted from the credential, i.e., to prove the consistency of the values with the commitment.
This statement proves that the revealed attributes are linked to their credential.
As a special case, if all attributes of a credential are specified as revealed in a policy, the client adds the full credential to the response instead of a proof.
For the policy rules on private attributes, the client computes a NIZK statement which proves that the attributes fulfill the rules.
Again, the client adds a statement to the proof that the attributes were indeed extracted from the credential.

For the computation of proofs, the client uses the common material retrieved in the setup phase and the NIZK statements.
The client also appends the metadata (e.g., issuer information) of all involved credentials to the response.

After executing the policy, the client encodes all proofs, revealed credentials, and credential metadata into a \emph{presentation token}.

Then, the user adds service-specific data and sends it alongside the token to the SP.

\textbf{(3) Verification of Presentation Tokens:}
On receiving a request, the SP loads the policy for the respective service.
The SP's policy interpreter then uses a \emph{NIZK verifier} and a \emph{policy verifier} to check the presentation token.

The SP extracts the NIZK proofs from the presentation token and verifies them using the NIZK verifier.
As inputs for the NIZK verifier, the policy and its proof system-specific representation, respectively, need to be provided.
Additionally, all public reference values, i.e., the commitments and all revealed attributes, need to be known to the NIZK verifier.
Hence, the SP extracts these values from the presentation token and provides them to the NIZK verifier.
After this step, the verifier is convinced that the proven statements match the requested statements.

As next step, the SP initializes the policy verifier and executes the remaining rules of the policy.
All rules that work only on revealed or public data are validated by the policy verifier.

\textbf{Ensuring Trustworthiness of Tokens:}
Besides evaluating the rules on the revealed attributes, this phase verifies the token's trustworthiness.
To do so, the policy verifier uses the trust rules encoded in the policy to check the issuer of the commitments.
That forms a trust chain from the issuer along the signature to the commitment, which is in turn linked with the proof and consequently the attributes.

The trust rules specify which issuers are trustworthy for which type of credentials.
That can, for instance, be done by providing a list of trusted issuers.
A more flexible method is to define a trusted trust scheme:
One example of a possible trust scheme is Europe's eIDAS trust framework.
Another example are SSI trust schemes established using distributed ledgers.
Depending on the defined trust scheme, the policy verifier automatically retrieves trust status information about the credential issuers from online registries.
This process ensures that public and private attributes can be trusted. Therefore, all NIZK statements on these attributes are trustworthy.%

After the NIZK verifier and the policy verifier conclude that the presentation token is trustworthy and fulfills the user's policy, the SP grants the user access to the service.

\begin{figure*}[h]
\centerline{ \includegraphics[]{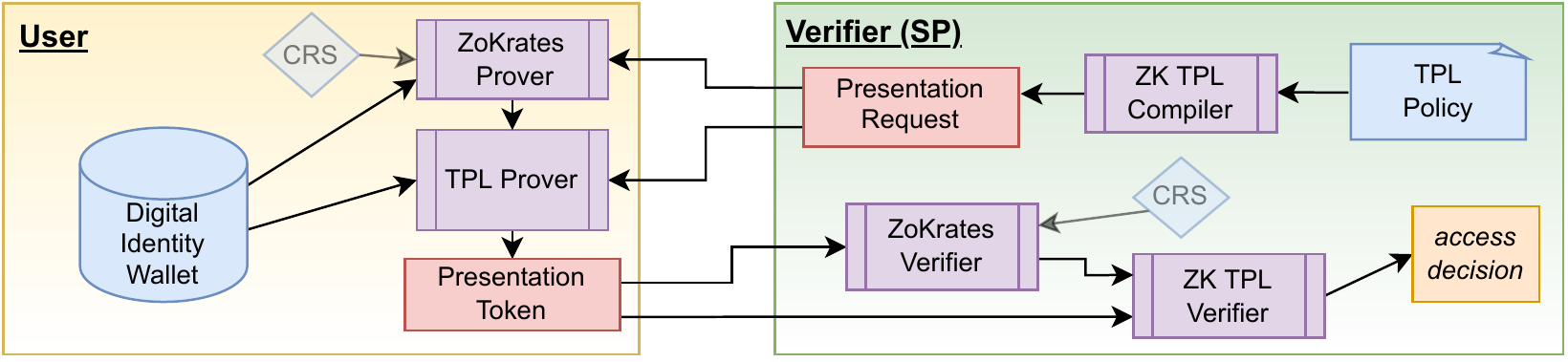}}
\caption{Architectural overview of our implementation including the actors and process flow.}\label{fig:arch}
\end{figure*}

\section{Implementation}
\label{sec:implementation}

While our concept is described on a generic level, the concrete choice of policy system and proof system is important to assess the feasibility and to evaluate the performance and security.
Thus, we provide a concrete instantiation of our concept.

Our implementation builds on the policy system \emph{TPL} (cf. \Cref{sec:background}), which we extend with privacy features.
Since TPL uses a logic-based syntax, we need to compile the rules encoded in form of TPL predicates to suitable statements for NIZK proofs.
As NIZK proof system we use SNARKs, since it enables small proofs.
We instantiate our SNARKs with the Groth16 proving scheme, which we execute with the help of the Bellman library.
Furthermore, we integrate the ZoKrates zero-knowledge toolbox~\cite{DBLP:conf/ithings/EberhardtT18} as an intermediate layer in the transformation process.
Thus, we compile TPL policies into ZoKrates programs, which are then mapped to circuits for bellman.
An advantage of this intermediate step is that the SP can already compile the policy into a ZoKrates program and directly share that with the user as part of the presentation request.
The user only needs to provide their credentials to the ZoKrates prover and execute that program.
Then, they send the resulting proof alongside the selected revealed credentials to the SP as part of the presentation token.
Finally, the SP uses the ZoKrates verifier to ensure the proof is valid.
In the next step, they forward the NIZK verification result and the rest of the presentation token to the TPL verifier.
In addition to executing the policy on the revealed credentials, the TPL verifier also checks whether all data is trustworthy.
An overview of this process is shown in \Cref{fig:arch}.

In our implementation we enable the privacy-preserving showing of attributes originating from credentials as well as private statements on these attributes driven by TPL policies.
In the following sections, we discuss the integration of our concept (cf. \Cref{sec:concept}) into the TPL system in more detail.
Specifically, \Cref{sec:extending-tpl} presents the extensions of TPL to ZK-TPL from the point of view of the policy author.
And \Cref{sec:compiling-tpl} covers aspects of compiling ZK-TPL into NIZK statements using ZoKrates.
Finally, \Cref{sec:eval} presents the evaluation of our implementation.

\subsection{Extending TPL with Zero-Knowledge Rules}
\label{sec:extending-tpl}

We now focus on the concrete changes to the TPL syntax to express ZK rules. %
A policy author defines in a policy which attributes a user needs to reveal.
There are multiple options to denote this in a TPL policy:
We now discuss a set of different options to extend the syntax TPL to do so.

\textbf{Option 1: Naming Convention:}
In TPL, the type of terms such as atoms and variables is defined by their name.
Any term starting with an uppercase letter followed by letters, numbers or underscores represents a variable.
Whereas terms starting with lowercase letters refer to atoms (constants).
Hence, in the same way we could refer to attributes that are not revealed via a naming convention.
However, as adding new conventions to the TPL specification would lead to ambiguities, we consider this approach to be error-prone and unintuitive.

\textbf{Option 2: Privacy Predicate:}
Another approach is to represent the hidden and revealed nature of attributes explicitly via a special predicate.
As with other domain-specific predicates that are available for TPL, a new predicate can be defined that only evaluates to true if the corresponding attribute is hidden.
With this approach, all predicates related to this attribute need then to cope with a potentially hidden attribute value.

\textbf{Option 3: \texttt{zkaccept}-Predicate:}
Finally, the third (and chosen) option is to add a new \texttt{zkaccept} predicate in addition to the \texttt{accept} entrypointy-predicate for TPL programs.
A policy is satisfied if and only if both \texttt{accept} and \texttt{zkaccept} evaluate to true.
Consequently, \texttt{accept} and \texttt{zkaccept} are the two main predicates that represent a TPL rule set.
With this approach, the meaning of the \texttt{accept} predicate is untouched and interpreted as before.
In the \texttt{zkaccept} predicates, all statements are interpreted with respect to hidden attributes and cause the creation and verification of the corresponding NIZK proofs.
Our approach is exemplified by the TPL policy in \Cref{tpl:zk-w3c}.
It requires the owner of the credential to be of 18 years or older, whereas neither the calculated age nor the \texttt{date\_{}of\_{}birth} attribute are revealed to the verifier.
The example policy also contains a \texttt{semester} attribute, which is revealed to the verifier since it is only used in the \texttt{accept} predicate, but not in the \texttt{zkaccept} predicate.

We opted to implement the third approach since it provides a clear differentiation between predicates applied to public and hidden data points.
As such, we consider it easier for the policy designer to design and reason about the policy.
From a technical perspective, we expect all three approaches to be implementable with reasonable effort.

\begin{listing}[h]
\begin{minted}[fontsize=\footnotesize, frame=single, framesep=5pt]{prolog}
zkaccept(Presentation) :-
 set_format(Presentation, w3cVP),
 extract(Presentation, verifiableCredential, Cred),
 set_format(Cred, w3cVC_student_id),
 extract(Cred, credentialSubject, Subject),

 extract(Subject, date_of_birth, Birthdate),
 calculateAge(Birthdate, Age), Age >= 18.

accept(Presentation) :-
 set_format(Presentation, w3cVP),
 extract(Presentation, verifiableCredential, Cred),
 set_format(Cred, w3cVC_student_id),
 extract(Cred, issuerDID, DIDissuer),
 checkQualified(DIDissuer),
 checkSig(Cred, DIDissuer),

 extract(Cred, credentialSubject, Subject),
 extract(Subject, semester, Semester),
 print(Semester).
\end{minted}
\caption{TPL policy from \Cref{tpl:basic-w3c} extended with privacy-preserving features: from the full credential, only the statement about age (derived from \texttt{Birthdate} and \texttt{Semester} attribute) are revealed.}
\label{tpl:zk-w3c}
\end{listing}

\textbf{Consistency Checks:}
When evaluating the policy on the prover or verifier side, one needs to take care of multiple issues.
First, an attribute can only appear as either hidden or revealed attribute.
If a revealed attribute is also used in the \texttt{zkaccept} predicate, this is likely a mistake of the policy designer.
Thus, the compiler needs to check if this invariant is satisfied and yield an alert if not. %
Secondly, when two or more distinct attributes of the same data structure, e.g., a credential, are used in \texttt{accept} and \texttt{zkaccept}, consistency needs to be ensured.
Hence, the zero-knowledge proof needs to be extended with statements ensuring consistency of all publicly revealed attributes.

\subsection{Compiling TPL policies to NIZK circuits}
\label{sec:compiling-tpl}

\textbf{ZK-TPL Compiler:}
As ZoKrates provides a domain-specific yet high-level language that closely resembles the syntax of Python, the TPL rules need to be compiled to this language.
ZoKrates itself then compiles the corresponding code to suitable circuits for the underlying NIZK library, i.e., bellman.
Hence, we provide the ZK-TPL compiler to transform policies from TPL syntax into ZoKrates' proof program syntax, as shown in \Cref{fig:arch}.
The ZoKrates standard library already provides several cryptographic primitives such as the compression function of SHA256, and SHA256 for a fixed number of calls to the compression functions, i.e., SHA256 for fixed input lengths, or Pedersen commitments.
Therefore, parts of the functionality we require are covered by the standard library.
Comparison operators for primitive types are also provided by ZoKrates.

When compiling TPL policies to ZoKrates programs, we consider the following challenges:

\textbf{1.) Constant-length Attributes:}
When generating the ZoKrates proof program, a challenge is to map attributes to either private or public variables, and how to encode their lengths.
As lengths can already be sensitive information for various data points, they are encoded as fixed-length string with \texttt{0} to pad to the maximal length.
Thus, there is a compromise between runtime costs for the additional padding, security, and functionality.
Length restrictions may be problematic for field types with international variations such as the use of first and last names.

\textbf{2.) Arithmetic:}
While integer types are available in various forms (8 bit to 64 bit), ZoKrates also provides a native \texttt{field} type representing $\mathbb{Z}_p$ where the prime $p$ depends on the choice of elliptic curve used by ZoKrates.
In general, $p$ will be large ($\geq 256$ bit) and all the arithmetic of the smaller types is implemented as $\mathbb{Z}_p$-arithmetic.
Hence, when compiling arithmetic involving hidden attributes, arithmetic is best represented using the \texttt{field} type unless specific features of the fixed bit-width types are needed.

\textbf{3.) Representing Strings as Numbers:}
Third, parsing arbitrary strings as integers is a complex and expensive task when performed inside ZoKrates.
Conversion of an array of \texttt{u8}s into a \texttt{field} involves arithmetic and potentially additional checks of well-formedness, e.g., that the individual bytes are ASCII digits, or that the full string is valid UTF-8.
Hence, we perform the parsing outside the ZK component as much as possible.
To ensure the integrity of the proof, this pre-processing step uses the same encoding of attributes than the issuer of the credential.
Thus, we require that the hash of the parsed data matches the hash used in the credential as commitment.

Note that our ZK-TPL compiler together with ZoKrates define the encoding of data and its representation in the rank-1 constraint system of the underlying SNARK. Therefore, any change our compiler or in ZoKrates may render old proofs unverifiable. For short-lived or interactive scenarios, we thus require compatible encodings for both prover and verifier.

\stefan{Elaborate more on how we did encoding/parsing, etc.}

\begin{listing}[t]
  \begin{minted}[fontsize=\footnotesize, frame=single, framesep=5pt,escapeinside=||]{python}
import "hashes/sha256/sha256" as sha256

def compare(u32[8] h1, u32[8] h2) -> bool:
  return h1[0] == h2[0] && h1[1] == h2[1] &&
         h1[2] == h2[2] && h1[3] == h2[3] &&
         h1[4] == h2[4] && h1[5] == h2[5] &&
         h1[6] == h2[6] && h1[7] == h2[7]

def main(u32[8] pub_hash, u32 currentYear,
    |\textbf{\underline{private}}| u32 birthYear, u32 semester) -> bool:

 |\textbf{// Encode full credential, append SHA256 padding:}|
 u32[1][16] enc_cred = [[birthYear, semester,
                     2147483648, 0, 0, 0, 0, 0,
                     0, 0, 0, 0, 0, 0, 0, 64]]
 |\textbf{// Calculate age using private attribute:}|
 u32 age = currentYear - birthYear
 |\textbf{// Proof that attributes fulfill the age check}|
 |\textbf{// and the consistency of data used for the proof:}|
 return age >= 18
     && compare(pub_hash, sha256(enc_cred))
  \end{minted}
  \caption{ZoKrates program generated by compiling the TPL policy from \Cref{tpl:zk-w3c}. Contains private \texttt{birthYear} attribute, revealed \texttt{semester} attribute, and (simplified) age check. It also proves the consistency of the attributes w.r.t. \texttt{pub\_{}hash} commitment of the credential. The magic-numbers for the encoded credential are SHA256 padding-constants. \label{lst:zokrates}}
\end{listing}

\textbf{Example ZoKrates Program:}
\Cref{lst:zokrates} presents an example ZK program, generated by our ZK-TPL compiler. %
For the inputs to the hash function, we opted to directly use \texttt{u32} arrays as expected by the provided implementation of SHA256.
When using different hash function designs with ZK-friendly permutations such as GMiMC~\cite{DBLP:conf/esorics/Albrecht0PRRR0S19} or Poseidon~\cite{DBLP:conf/uss/0001KR0S21}, the inputs and outputs can be represented as \texttt{field} elements.
Thereby, we would be able to significantly improve the performance of the ZK program.
With the goal in mind to be as widely usable as possible, we consider support for common hashes such as SHA256 essential.
\label{sec:impl:shaPoseidon}

\begin{figure}[htbp]
\centerline{\includegraphics[width=\columnwidth]{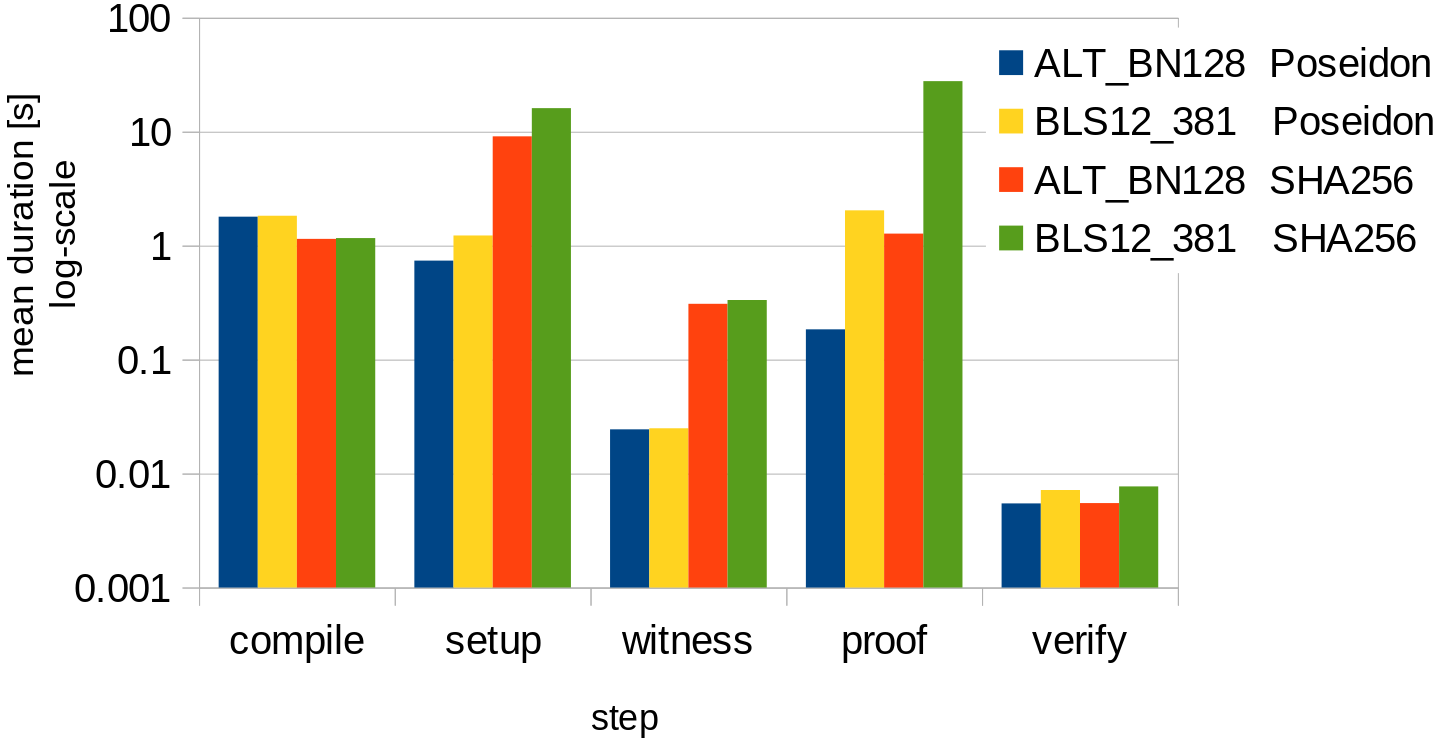}}
\caption[]{\label{fig:bench1} Overhead evaluation results of the example policy in \Cref{tpl:zk-w3c} for different commitments and curves.}
\end{figure}

\subsection{Evaluation}
\label{sec:eval}

To evaluate our proof-of-concept implementation
perform several benchmarks and compare them with the evaluation of the TPL system without any privacy features.
Existing TPL benchmarks focus on the verification phase, which takes one to ten seconds for realistic policies and includes the retrieval of trust information from online registries~\cite{DBLP:conf/IEEEares/MoreA22}.
Standard TPL needs no setup phase, and the authentication phase is a trivial process for the user.

\textbf{Setup:}
To measure the impact of the privacy extension to the existing TPL toolchain in Java,
we run benchmarks on a 2022 business laptop. %
Our prover and verifier tools use ZoKrates, which we configured to use the bellmann backend with Groth16~\cite{DBLP:conf/eurocrypt/Groth16}.
We evaluate the performance on the \texttt{ALT\_{}BN128}~\cite{DBLP:conf/sacrypt/BarretoN05} and \texttt{BLS12\_{}381}~\cite{BLS12-381} curves; while the former provides $100$ bit of security, the latter is slower but provides $\geq117$ bit of security~\cite{DBLP:journals/joc/BarbulescuD19,draft-yonezawa-pairing-friendly-curves-02}.
Additionally, we compare the performance of SHA256 with ZK-friendly Poseidon~\cite{DBLP:conf/uss/0001KR0S21} (cf.~\Cref{sec:impl:shaPoseidon}).

\textbf{Results:}
We divide our three phases (cf.~\Cref{sec:concept}) in two categories:
\begin{enumerate*}
  \item The setup phase (compilation of policy, CRS setup) is a one-time phase and only performed once for each policy.
  \item The authentication phase (computation of witness, generation of proof) and verification phase (verification of proof and execution of policy) are repeated phases and executed for each authentication process.
\end{enumerate*}

\Cref{fig:bench1} visualizes the results of our evaluation.
We observe that %
the verification duration and the size of the proof transmitted to the verifier are independent of the complexity of the policy.
This is because the size of SNARK proofs amounts to only 3 elliptic curve points and is independent of the witness size.
Also, the size of the transmitted proof naturally depends on the size of the revealed attributes.
The performance of the setup and authentication phases depend on the number of attributes that are part of the credential.
This is because all attributes are part of the credential's signature and thus need to be part of the commitment in the proof (cf.~\Cref{lst:zokrates}).
The performance of calculating the commitment in the proof program is linear in the size of the attributes.

\section{Discussion}
\label{sec:discussion}

\textbf{NIZK Setup:}
From a deployment point of view, policy-dependent setup phases may hamper the adoption of such a system.
As this dependency is mainly influenced by the underlying proof system, an efficient proof system with a universal CRS is of paramount importance for more flexible applications.
Also, the need for a trusted third party for the CRS generation is not ideal in some use cases.
However, it has already been shown that the TTP can be replaced using secure multi-party computation for the setup algorithm~\cite{DBLP:conf/fc/BoweGG18}.
Alternatively, it is also possible to employ transparent, subversion-resistant, or updatable proof systems~\cite{DBLP:conf/crypto/Ben-SassonBHR19,DBLP:conf/eurocrypt/BunzFS20,DBLP:conf/pkc/Fuchsbauer18,DBLP:conf/crypto/GrothKMMM18,DBLP:conf/ccs/AbdolmalekiRS20}, where knowledge of secret trapdoors no longer poses a threat.

\textbf{Constant-length Attributes:}
During the implementation of the ZK-TPL interpreter and the design of example ZK-TPL policies,
we observe multiple restrictions inherent to the use of attributes with arbitrary types.
Specifically, when dealing with string attributes, all the strings need to be encoded with a constant length.
Otherwise, the length of the strings could reduce the anonymity set and the mere knowledge of the string lengths leaks sensitive information.
This also extends to primitives that consume these strings, e.g., hash functions, as the number of compression function evaluations depends on the size of the input.

\textbf{Future Work on NIZK Toolchains:}
While NIZKs and SNARKs are known for languages in all of NP (cf. \cite{DBLP:conf/eurocrypt/GrothOS06} and others), for practical purposes the situation is significantly different.
Yet, as implementations of SNARKs gain better toolchains with support for higher-level abstractions, SNARKs can be applied to solve more interesting challenges.
These toolchains need to abstract technical details such as rank-1 constraint systems and other arithmetization techniques to be useful for a wider audience.
With ongoing scientific and engineering effort, these abstractions are rendered more efficient, less costly, and more expressive.

\textbf{Future Work on Policy Authoring Tools:}
Since the capabilities of the policy language got extended, we need to update the GUI-based authoring tools~\cite{DBLP:conf/openidentity/ModersheimN19,DBLP:conf/icissp/WeinhardtO19} in future work.
Attributes should be hidden by default and only be revealed when indicated.
Non-technical policy authors should be able to use zero-knowledge features in a graphical manner without being familiar any underlying details.

\textbf{Communicating Privacy Implications to Users:}
While we extend the capabilities available to a policy designer, the consequences of revealing certain attributes also need to be explained to the user.
Some works have developed interfaces that highlight the revealed attributes and data flows to the user.
Examples in various directions include Angulo et al.'s
approach~\cite{DBLP:books/sp/primelife11/AnguloFPK11}, which provides visualizations of policies, and Mikkelsen et al.~\cite{DBLP:books/daglib/p/MikkelsenDGJGNPPSSZ15} presenting a user interface to disclose attributes of a credential selectively.
Alternatively, privacy metrics~\cite{DBLP:journals/csur/WagnerE18} offer tools to attach scores based on the processed data and the type of performed computations.
Using these techniques may help visualize a user's potential privacy risks based on the policy in question.

\begin{acks}
This work was supported by the
\grantsponsor{EUH2020}{European Union's Horizon 2020 research and innovation programme}{https://ec.europa.eu/programmes/horizon2020/en} under grant agreement \textnumero~\grantnum{EUH2020}{871473} (KRAKEN).
We thank our colleague Jakob Heher for proofreading our manuscript.
\end{acks}

\bibliographystyle{ACM-Reference-Format}
\bibliography{bib,dblp}


\begin{thebibliography}{51}

%%% the Web address.

\ifx \showCODEN    \undefined \def \showCODEN     #1{\unskip}     \fi
\ifx \showDOI      \undefined \def \showDOI       #1{#1}\fi
\ifx \showISBNx    \undefined \def \showISBNx     #1{\unskip}     \fi
\ifx \showISBNxiii \undefined \def \showISBNxiii  #1{\unskip}     \fi
\ifx \showISSN     \undefined \def \showISSN      #1{\unskip}     \fi
\ifx \showLCCN     \undefined \def \showLCCN      #1{\unskip}     \fi
\ifx \shownote     \undefined \def \shownote      #1{#1}          \fi
\ifx \showarticletitle \undefined \def \showarticletitle #1{#1}   \fi
\ifx \showURL      \undefined \def \showURL       {\relax}        \fi
% invisible to TeX
\providecommand\bibfield[2]{#2}
\providecommand\bibinfo[2]{#2}
\providecommand\natexlab[1]{#1}
\providecommand\showeprint[2][]{arXiv:#2}

\bibitem[\protect\citeauthoryear{Abdolmaleki, Ramacher, and
  Slamanig}{Abdolmaleki et~al\mbox{.}}{2020}]%
        {DBLP:conf/ccs/AbdolmalekiRS20}
\bibfield{author}{\bibinfo{person}{Behzad Abdolmaleki},
  \bibinfo{person}{Sebastian Ramacher}, {and} \bibinfo{person}{Daniel
  Slamanig}.} \bibinfo{year}{2020}\natexlab{}.
\newblock \showarticletitle{Lift-and-Shift: Obtaining Simulation Extractable
  Subversion and Updatable SNARKs Generically}. In
  \bibinfo{booktitle}{\emph{{CCS}}}. \bibinfo{publisher}{{ACM}},
  \bibinfo{pages}{1987--2005}.
\newblock


\bibitem[\protect\citeauthoryear{Abraham, H{\"{o}}randner, Omolola, and
  Ramacher}{Abraham et~al\mbox{.}}{2019}]%
        {DBLP:conf/icics/AbrahamHOR19}
\bibfield{author}{\bibinfo{person}{Andreas Abraham}, \bibinfo{person}{Felix
  H{\"{o}}randner}, \bibinfo{person}{Olamide Omolola}, {and}
  \bibinfo{person}{Sebastian Ramacher}.} \bibinfo{year}{2019}\natexlab{}.
\newblock \showarticletitle{Privacy-Preserving eID Derivation for
  Self-Sovereign Identity Systems}. In \bibinfo{booktitle}{\emph{{ICICS}}}
  \emph{(\bibinfo{series}{{LNCS}}, Vol.~\bibinfo{volume}{11999})}.
  \bibinfo{publisher}{Springer}, \bibinfo{pages}{307--323}.
\newblock


\bibitem[\protect\citeauthoryear{Abraham, Koch, More, Ramacher, and
  Stopar}{Abraham et~al\mbox{.}}{2021}]%
        {DBLP:conf/trustcom/AbrahamKMRS21}
\bibfield{author}{\bibinfo{person}{Andreas Abraham}, \bibinfo{person}{Karl
  Koch}, \bibinfo{person}{Stefan More}, \bibinfo{person}{Sebastian Ramacher},
  {and} \bibinfo{person}{Miha Stopar}.} \bibinfo{year}{2021}\natexlab{}.
\newblock \showarticletitle{Privacy-Preserving eID Derivation to Self-Sovereign
  Identity Systems with Offline Revocation}. In
  \bibinfo{booktitle}{\emph{TrustCom}}. \bibinfo{publisher}{{IEEE}},
  \bibinfo{pages}{506--513}.
\newblock


\bibitem[\protect\citeauthoryear{Alber, More, M{\"{o}}dersheim, and
  Schlichtkrull}{Alber et~al\mbox{.}}{2021}]%
        {DBLP:conf/openidentity/AlberMMS21}
\bibfield{author}{\bibinfo{person}{Lukas Alber}, \bibinfo{person}{Stefan More},
  \bibinfo{person}{Sebastian M{\"{o}}dersheim}, {and} \bibinfo{person}{Anders
  Schlichtkrull}.} \bibinfo{year}{2021}\natexlab{}.
\newblock \showarticletitle{Adapting the {TPL} Trust Policy Language for a
  Self-Sovereign Identity World}. In \bibinfo{booktitle}{\emph{Open Identity
  Summit}} \emph{(\bibinfo{series}{{LNI}}, Vol.~\bibinfo{volume}{{P-312}})}.
  \bibinfo{publisher}{Gesellschaft f{\"{u}}r Informatik e.V.},
  \bibinfo{pages}{107--118}.
\newblock


\bibitem[\protect\citeauthoryear{Albrecht, Grassi, Perrin, Ramacher,
  Rechberger, Rotaru, Roy, and Schofnegger}{Albrecht et~al\mbox{.}}{2019}]%
        {DBLP:conf/esorics/Albrecht0PRRR0S19}
\bibfield{author}{\bibinfo{person}{Martin~R. Albrecht},
  \bibinfo{person}{Lorenzo Grassi}, \bibinfo{person}{L{\'{e}}o Perrin},
  \bibinfo{person}{Sebastian Ramacher}, \bibinfo{person}{Christian Rechberger},
  \bibinfo{person}{Dragos Rotaru}, \bibinfo{person}{Arnab Roy}, {and}
  \bibinfo{person}{Markus Schofnegger}.} \bibinfo{year}{2019}\natexlab{}.
\newblock \showarticletitle{Feistel Structures for MPC, and More}. In
  \bibinfo{booktitle}{\emph{{ESORICS} {(2)}}} \emph{(\bibinfo{series}{{LNCS}},
  Vol.~\bibinfo{volume}{11736})}. \bibinfo{publisher}{Springer},
  \bibinfo{pages}{151--171}.
\newblock


\bibitem[\protect\citeauthoryear{Allen}{Allen}{2016}]%
        {allen2016}
\bibfield{author}{\bibinfo{person}{Christopher Allen}.}
  \bibinfo{year}{2016}\natexlab{}.
\newblock \bibinfo{title}{{The Path to Self-Sovereign-Identity}}.
\newblock
\newblock
\urldef\tempurl%
\url{http://www.lifewithalacrity.com/2016/04/the-path-to-self-soverereign-identity.html}
\showURL{%
\tempurl}
\newblock
\shownote{{online, accessed on 2022-07-13}.}


\bibitem[\protect\citeauthoryear{Alm, Wolf, and Posegga}{Alm
  et~al\mbox{.}}{2009}]%
        {DBLP:conf/trustbus/AlmWP09}
\bibfield{author}{\bibinfo{person}{Christopher Alm}, \bibinfo{person}{Ruben
  Wolf}, {and} \bibinfo{person}{Joachim Posegga}.}
  \bibinfo{year}{2009}\natexlab{}.
\newblock \showarticletitle{The {OPL} Access Control Policy Language}. In
  \bibinfo{booktitle}{\emph{TrustBus}} \emph{(\bibinfo{series}{{LNCS}},
  Vol.~\bibinfo{volume}{5695})}. \bibinfo{publisher}{Springer},
  \bibinfo{pages}{138--148}.
\newblock


\bibitem[\protect\citeauthoryear{Angulo, Fischer{-}H{\"{u}}bner, Pulls, and
  K{\"{o}}nig}{Angulo et~al\mbox{.}}{2011}]%
        {DBLP:books/sp/primelife11/AnguloFPK11}
\bibfield{author}{\bibinfo{person}{Julio Angulo}, \bibinfo{person}{Simone
  Fischer{-}H{\"{u}}bner}, \bibinfo{person}{Tobias Pulls}, {and}
  \bibinfo{person}{Ulrich K{\"{o}}nig}.} \bibinfo{year}{2011}\natexlab{}.
\newblock \showarticletitle{{HCI} for Policy Display and Administration}.
\newblock In \bibinfo{booktitle}{\emph{Privacy and Identity Management for
  Life}}. \bibinfo{publisher}{Springer}, \bibinfo{pages}{261--277}.
\newblock


\bibitem[\protect\citeauthoryear{Barbulescu and Duquesne}{Barbulescu and
  Duquesne}{2019}]%
        {DBLP:journals/joc/BarbulescuD19}
\bibfield{author}{\bibinfo{person}{Razvan Barbulescu} {and}
  \bibinfo{person}{Sylvain Duquesne}.} \bibinfo{year}{2019}\natexlab{}.
\newblock \showarticletitle{Updating Key Size Estimations for Pairings}.
\newblock \bibinfo{journal}{\emph{J. Cryptol.}} \bibinfo{volume}{32},
  \bibinfo{number}{4} (\bibinfo{year}{2019}), \bibinfo{pages}{1298--1336}.
\newblock


\bibitem[\protect\citeauthoryear{Barreto and Naehrig}{Barreto and
  Naehrig}{2005}]%
        {DBLP:conf/sacrypt/BarretoN05}
\bibfield{author}{\bibinfo{person}{Paulo S. L.~M. Barreto} {and}
  \bibinfo{person}{Michael Naehrig}.} \bibinfo{year}{2005}\natexlab{}.
\newblock \showarticletitle{Pairing-Friendly Elliptic Curves of Prime Order}.
  In \bibinfo{booktitle}{\emph{{SAC}}} \emph{(\bibinfo{series}{{LNCS}},
  Vol.~\bibinfo{volume}{3897})}. \bibinfo{publisher}{Springer},
  \bibinfo{pages}{319--331}.
\newblock


\bibitem[\protect\citeauthoryear{Belchior, Putz, Pernul, Correia, Vasconcelos,
  and Guerreiro}{Belchior et~al\mbox{.}}{2020}]%
        {DBLP:conf/trustcom/BelchiorPP0VG20}
\bibfield{author}{\bibinfo{person}{Rafael Belchior}, \bibinfo{person}{Benedikt
  Putz}, \bibinfo{person}{G{\"{u}}nther Pernul}, \bibinfo{person}{Miguel
  Correia}, \bibinfo{person}{Andr{\'{e}} Vasconcelos}, {and}
  \bibinfo{person}{S{\'{e}}rgio Guerreiro}.} \bibinfo{year}{2020}\natexlab{}.
\newblock \showarticletitle{{SSIBAC:} Self-Sovereign Identity Based Access
  Control}. In \bibinfo{booktitle}{\emph{TrustCom}}.
  \bibinfo{publisher}{{IEEE}}, \bibinfo{pages}{1935--1943}.
\newblock


\bibitem[\protect\citeauthoryear{Ben{-}Sasson, Bentov, Horesh, and
  Riabzev}{Ben{-}Sasson et~al\mbox{.}}{2019}]%
        {DBLP:conf/crypto/Ben-SassonBHR19}
\bibfield{author}{\bibinfo{person}{Eli Ben{-}Sasson}, \bibinfo{person}{Iddo
  Bentov}, \bibinfo{person}{Yinon Horesh}, {and} \bibinfo{person}{Michael
  Riabzev}.} \bibinfo{year}{2019}\natexlab{}.
\newblock \showarticletitle{Scalable Zero Knowledge with No Trusted Setup}. In
  \bibinfo{booktitle}{\emph{{CRYPTO} {(3)}}} \emph{(\bibinfo{series}{{LNCS}},
  Vol.~\bibinfo{volume}{11694})}. \bibinfo{publisher}{Springer},
  \bibinfo{pages}{701--732}.
\newblock


\bibitem[\protect\citeauthoryear{Bichsel, Camenisch, Dubovitskaya, Enderlein,
  Krenn, Krontiris, Lehmann, Neven, Paquin, Preiss, Rannenberg, and
  Sabouri}{Bichsel et~al\mbox{.}}{2015a}]%
        {DBLP:books/daglib/p/BichselCDEKKLNPPRS15}
\bibfield{author}{\bibinfo{person}{Patrik Bichsel}, \bibinfo{person}{Jan
  Camenisch}, \bibinfo{person}{Maria Dubovitskaya}, \bibinfo{person}{Robert~R.
  Enderlein}, \bibinfo{person}{Stephan Krenn}, \bibinfo{person}{Ioannis
  Krontiris}, \bibinfo{person}{Anja Lehmann}, \bibinfo{person}{Gregory Neven},
  \bibinfo{person}{Christian Paquin}, \bibinfo{person}{Franz{-}Stefan Preiss},
  \bibinfo{person}{Kai Rannenberg}, {and} \bibinfo{person}{Ahmad Sabouri}.}
  \bibinfo{year}{2015}\natexlab{a}.
\newblock \showarticletitle{An Architecture for Privacy-ABCs}.
\newblock In \bibinfo{booktitle}{\emph{Attribute-based Credentials for Trust}}.
  \bibinfo{publisher}{Springer}, \bibinfo{pages}{11--78}.
\newblock


\bibitem[\protect\citeauthoryear{Bichsel, Camenisch, Dubovitskaya, Enderlein,
  Krenn, Lehmann, Neven, and Preiss}{Bichsel et~al\mbox{.}}{2015b}]%
        {DBLP:books/daglib/p/BichselCDEKLNP15}
\bibfield{author}{\bibinfo{person}{Patrik Bichsel}, \bibinfo{person}{Jan
  Camenisch}, \bibinfo{person}{Maria Dubovitskaya}, \bibinfo{person}{Robert~R.
  Enderlein}, \bibinfo{person}{Stephan Krenn}, \bibinfo{person}{Anja Lehmann},
  \bibinfo{person}{Gregory Neven}, {and} \bibinfo{person}{Franz{-}Stefan
  Preiss}.} \bibinfo{year}{2015}\natexlab{b}.
\newblock \showarticletitle{Cryptographic Protocols Underlying Privacy-ABCs}.
\newblock In \bibinfo{booktitle}{\emph{Attribute-based Credentials for Trust}}.
  \bibinfo{publisher}{Springer}, \bibinfo{pages}{79--108}.
\newblock


\bibitem[\protect\citeauthoryear{Bitansky, Canetti, Chiesa, and
  Tromer}{Bitansky et~al\mbox{.}}{2012}]%
        {DBLP:conf/innovations/BitanskyCCT12}
\bibfield{author}{\bibinfo{person}{Nir Bitansky}, \bibinfo{person}{Ran
  Canetti}, \bibinfo{person}{Alessandro Chiesa}, {and} \bibinfo{person}{Eran
  Tromer}.} \bibinfo{year}{2012}\natexlab{}.
\newblock \showarticletitle{From extractable collision resistance to succinct
  non-interactive arguments of knowledge, and back again}. In
  \bibinfo{booktitle}{\emph{{ITCS}}}. \bibinfo{publisher}{{ACM}},
  \bibinfo{pages}{326--349}.
\newblock


\bibitem[\protect\citeauthoryear{Blaze, Feigenbaum, and Lacy}{Blaze
  et~al\mbox{.}}{1996}]%
        {DBLP:conf/sp/BlazeFL96}
\bibfield{author}{\bibinfo{person}{Matt Blaze}, \bibinfo{person}{Joan
  Feigenbaum}, {and} \bibinfo{person}{Jack Lacy}.}
  \bibinfo{year}{1996}\natexlab{}.
\newblock \showarticletitle{Decentralized Trust Management}. In
  \bibinfo{booktitle}{\emph{{IEEE} S\&P}}. \bibinfo{publisher}{{IEEE}},
  \bibinfo{pages}{164--173}.
\newblock


\bibitem[\protect\citeauthoryear{Bowe}{Bowe}{2017}]%
        {BLS12-381}
\bibfield{author}{\bibinfo{person}{Sean Bowe}.}
  \bibinfo{year}{2017}\natexlab{}.
\newblock \bibinfo{booktitle}{\emph{\texttt{BLS12-381:} New zk-SNARK elliptic
  curve construction}}.
\newblock
\urldef\tempurl%
\url{https://electriccoin.co/blog/new-snark-curve/}
\showURL{%
\tempurl}


\bibitem[\protect\citeauthoryear{Bowe, Gabizon, and Green}{Bowe
  et~al\mbox{.}}{2018}]%
        {DBLP:conf/fc/BoweGG18}
\bibfield{author}{\bibinfo{person}{Sean Bowe}, \bibinfo{person}{Ariel Gabizon},
  {and} \bibinfo{person}{Matthew~D. Green}.} \bibinfo{year}{2018}\natexlab{}.
\newblock \showarticletitle{A Multi-party Protocol for Constructing the Public
  Parameters of the Pinocchio zk-SNARK}. In \bibinfo{booktitle}{\emph{Financial
  Cryptography Workshops}} \emph{(\bibinfo{series}{{LNCS}},
  Vol.~\bibinfo{volume}{10958})}. \bibinfo{publisher}{Springer},
  \bibinfo{pages}{64--77}.
\newblock


\bibitem[\protect\citeauthoryear{Bruegger and Lipp}{Bruegger and Lipp}{2016}]%
        {DBLP:conf/openidentity/BrueggerL16}
\bibfield{author}{\bibinfo{person}{Bud~P. Bruegger} {and}
  \bibinfo{person}{Peter Lipp}.} \bibinfo{year}{2016}\natexlab{}.
\newblock \showarticletitle{LIGHT\({}^{\mbox{est}}\) - {A} Lightweight
  Infrastructure for Global Heterogeneous Trust Management}. In
  \bibinfo{booktitle}{\emph{Open Identity Summit}}
  \emph{(\bibinfo{series}{{LNI}}, Vol.~\bibinfo{volume}{{P-264}})}.
  \bibinfo{publisher}{{GI}}, \bibinfo{pages}{15--26}.
\newblock


\bibitem[\protect\citeauthoryear{B{\"{u}}nz, Fisch, and Szepieniec}{B{\"{u}}nz
  et~al\mbox{.}}{2020}]%
        {DBLP:conf/eurocrypt/BunzFS20}
\bibfield{author}{\bibinfo{person}{Benedikt B{\"{u}}nz}, \bibinfo{person}{Ben
  Fisch}, {and} \bibinfo{person}{Alan Szepieniec}.}
  \bibinfo{year}{2020}\natexlab{}.
\newblock \showarticletitle{Transparent SNARKs from {DARK} Compilers}. In
  \bibinfo{booktitle}{\emph{{EUROCRYPT} {(1)}}}
  \emph{(\bibinfo{series}{{LNCS}}, Vol.~\bibinfo{volume}{12105})}.
  \bibinfo{publisher}{Springer}, \bibinfo{pages}{677--706}.
\newblock


\bibitem[\protect\citeauthoryear{Camenisch and Lysyanskaya}{Camenisch and
  Lysyanskaya}{2002}]%
        {DBLP:conf/scn/CamenischL02}
\bibfield{author}{\bibinfo{person}{Jan Camenisch} {and} \bibinfo{person}{Anna
  Lysyanskaya}.} \bibinfo{year}{2002}\natexlab{}.
\newblock \showarticletitle{A Signature Scheme with Efficient Protocols}. In
  \bibinfo{booktitle}{\emph{{SCN}}} \emph{(\bibinfo{series}{{LNCS}},
  Vol.~\bibinfo{volume}{2576})}. \bibinfo{publisher}{Springer},
  \bibinfo{pages}{268--289}.
\newblock


\bibitem[\protect\citeauthoryear{Camenisch, Mödersheim, Neven, Preiss, and
  Rial}{Camenisch et~al\mbox{.}}{2015}]%
        {prologABCmatcher}
\bibfield{author}{\bibinfo{person}{Jan Camenisch}, \bibinfo{person}{Sebastian
  Mödersheim}, \bibinfo{person}{Gregory Neven}, \bibinfo{person}{Franz-Stefan
  Preiss}, {and} \bibinfo{person}{Alfredo Rial}.}
  \bibinfo{year}{2015}\natexlab{}.
\newblock \bibinfo{booktitle}{\emph{A Prolog Program for Matching
  Attribute-Based Credentials to Access Control Policies}}.
\newblock \bibinfo{type}{Research Report}. \bibinfo{institution}{IBM}.
\newblock


\bibitem[\protect\citeauthoryear{Chaum}{Chaum}{1981}]%
        {DBLP:journals/cacm/Chaum81}
\bibfield{author}{\bibinfo{person}{David Chaum}.}
  \bibinfo{year}{1981}\natexlab{}.
\newblock \showarticletitle{Untraceable Electronic Mail, Return Addresses, and
  Digital Pseudonyms}.
\newblock \bibinfo{journal}{\emph{Commun. {ACM}}} \bibinfo{volume}{24},
  \bibinfo{number}{2} (\bibinfo{year}{1981}), \bibinfo{pages}{84--88}.
\newblock


\bibitem[\protect\citeauthoryear{Chaum}{Chaum}{1985}]%
        {DBLP:journals/cacm/Chaum85}
\bibfield{author}{\bibinfo{person}{David Chaum}.}
  \bibinfo{year}{1985}\natexlab{}.
\newblock \showarticletitle{Security Without Identification: Transaction
  Systems to Make Big Brother Obsolete}.
\newblock \bibinfo{journal}{\emph{Commun. {ACM}}} \bibinfo{volume}{28},
  \bibinfo{number}{10} (\bibinfo{year}{1985}), \bibinfo{pages}{1030--1044}.
\newblock


\bibitem[\protect\citeauthoryear{Coi and Olmedilla}{Coi and Olmedilla}{2008}]%
        {DBLP:conf/secrypt/CoiO08}
\bibfield{author}{\bibinfo{person}{Juri Luca~De Coi} {and}
  \bibinfo{person}{Daniel Olmedilla}.} \bibinfo{year}{2008}\natexlab{}.
\newblock \showarticletitle{A Review of Trust Management, Security and Privacy
  Policy Languages}. In \bibinfo{booktitle}{\emph{{SECRYPT}}}.
  \bibinfo{publisher}{{INSTICC} Press}, \bibinfo{pages}{483--490}.
\newblock


\bibitem[\protect\citeauthoryear{Eberhardt and Tai}{Eberhardt and Tai}{2018}]%
        {DBLP:conf/ithings/EberhardtT18}
\bibfield{author}{\bibinfo{person}{Jacob Eberhardt} {and}
  \bibinfo{person}{Stefan Tai}.} \bibinfo{year}{2018}\natexlab{}.
\newblock \showarticletitle{ZoKrates - Scalable Privacy-Preserving Off-Chain
  Computations}. In
  \bibinfo{booktitle}{\emph{iThings/GreenCom/CPSCom/SmartData}}.
  \bibinfo{publisher}{{IEEE}}, \bibinfo{pages}{1084--1091}.
\newblock


\bibitem[\protect\citeauthoryear{Fuchsbauer}{Fuchsbauer}{2018}]%
        {DBLP:conf/pkc/Fuchsbauer18}
\bibfield{author}{\bibinfo{person}{Georg Fuchsbauer}.}
  \bibinfo{year}{2018}\natexlab{}.
\newblock \showarticletitle{Subversion-Zero-Knowledge SNARKs}. In
  \bibinfo{booktitle}{\emph{{PKC} {(1)}}} \emph{(\bibinfo{series}{{LNCS}},
  Vol.~\bibinfo{volume}{10769})}. \bibinfo{publisher}{Springer},
  \bibinfo{pages}{315--347}.
\newblock


\bibitem[\protect\citeauthoryear{Fuchsbauer, Hanser, and Slamanig}{Fuchsbauer
  et~al\mbox{.}}{2019}]%
        {DBLP:journals/joc/FuchsbauerHS19}
\bibfield{author}{\bibinfo{person}{Georg Fuchsbauer},
  \bibinfo{person}{Christian Hanser}, {and} \bibinfo{person}{Daniel Slamanig}.}
  \bibinfo{year}{2019}\natexlab{}.
\newblock \showarticletitle{Structure-Preserving Signatures on Equivalence
  Classes and Constant-Size Anonymous Credentials}.
\newblock \bibinfo{journal}{\emph{J. Cryptol.}} \bibinfo{volume}{32},
  \bibinfo{number}{2} (\bibinfo{year}{2019}), \bibinfo{pages}{498--546}.
\newblock


\bibitem[\protect\citeauthoryear{Grassi, Khovratovich, Rechberger, Roy, and
  Schofnegger}{Grassi et~al\mbox{.}}{2021}]%
        {DBLP:conf/uss/0001KR0S21}
\bibfield{author}{\bibinfo{person}{Lorenzo Grassi}, \bibinfo{person}{Dmitry
  Khovratovich}, \bibinfo{person}{Christian Rechberger}, \bibinfo{person}{Arnab
  Roy}, {and} \bibinfo{person}{Markus Schofnegger}.}
  \bibinfo{year}{2021}\natexlab{}.
\newblock \showarticletitle{Poseidon: {A} New Hash Function for Zero-Knowledge
  Proof Systems}. In \bibinfo{booktitle}{\emph{{USENIX}}}.
  \bibinfo{publisher}{{USENIX} Association}, \bibinfo{pages}{519--535}.
\newblock


\bibitem[\protect\citeauthoryear{Groth}{Groth}{2016}]%
        {DBLP:conf/eurocrypt/Groth16}
\bibfield{author}{\bibinfo{person}{Jens Groth}.}
  \bibinfo{year}{2016}\natexlab{}.
\newblock \showarticletitle{On the Size of Pairing-Based Non-interactive
  Arguments}. In \bibinfo{booktitle}{\emph{{EUROCRYPT} {(2)}}}
  \emph{(\bibinfo{series}{{LNCS}}, Vol.~\bibinfo{volume}{9666})}.
  \bibinfo{publisher}{Springer}, \bibinfo{pages}{305--326}.
\newblock


\bibitem[\protect\citeauthoryear{Groth, Kohlweiss, Maller, Meiklejohn, and
  Miers}{Groth et~al\mbox{.}}{2018}]%
        {DBLP:conf/crypto/GrothKMMM18}
\bibfield{author}{\bibinfo{person}{Jens Groth}, \bibinfo{person}{Markulf
  Kohlweiss}, \bibinfo{person}{Mary Maller}, \bibinfo{person}{Sarah
  Meiklejohn}, {and} \bibinfo{person}{Ian Miers}.}
  \bibinfo{year}{2018}\natexlab{}.
\newblock \showarticletitle{Updatable and Universal Common Reference Strings
  with Applications to zk-SNARKs}. In \bibinfo{booktitle}{\emph{{CRYPTO}
  {(3)}}} \emph{(\bibinfo{series}{{LNCS}}, Vol.~\bibinfo{volume}{10993})}.
  \bibinfo{publisher}{Springer}, \bibinfo{pages}{698--728}.
\newblock


\bibitem[\protect\citeauthoryear{Groth, Ostrovsky, and Sahai}{Groth
  et~al\mbox{.}}{2006}]%
        {DBLP:conf/eurocrypt/GrothOS06}
\bibfield{author}{\bibinfo{person}{Jens Groth}, \bibinfo{person}{Rafail
  Ostrovsky}, {and} \bibinfo{person}{Amit Sahai}.}
  \bibinfo{year}{2006}\natexlab{}.
\newblock \showarticletitle{Perfect Non-interactive Zero Knowledge for {NP}}.
  In \bibinfo{booktitle}{\emph{{EUROCRYPT}}} \emph{(\bibinfo{series}{{LNCS}},
  Vol.~\bibinfo{volume}{4004})}. \bibinfo{publisher}{Springer},
  \bibinfo{pages}{339--358}.
\newblock


\bibitem[\protect\citeauthoryear{Koch, Krenn, Pellegrino, and Ramacher}{Koch
  et~al\mbox{.}}{2020}]%
        {DBLP:conf/primelife/KochKPR20}
\bibfield{author}{\bibinfo{person}{Karl Koch}, \bibinfo{person}{Stephan Krenn},
  \bibinfo{person}{Donato Pellegrino}, {and} \bibinfo{person}{Sebastian
  Ramacher}.} \bibinfo{year}{2020}\natexlab{}.
\newblock \showarticletitle{Privacy-Preserving Analytics for Data Markets Using
  {MPC}}. In \bibinfo{booktitle}{\emph{Privacy and Identity Management}}
  \emph{(\bibinfo{series}{{IFIP} Advances in Information and Communication
  Technology}, Vol.~\bibinfo{volume}{619})}. \bibinfo{publisher}{Springer},
  \bibinfo{pages}{226--246}.
\newblock


\bibitem[\protect\citeauthoryear{Kosba, Papamanthou, and Shi}{Kosba
  et~al\mbox{.}}{2018}]%
        {DBLP:conf/sp/KosbaPS18}
\bibfield{author}{\bibinfo{person}{Ahmed~E. Kosba},
  \bibinfo{person}{Charalampos Papamanthou}, {and} \bibinfo{person}{Elaine
  Shi}.} \bibinfo{year}{2018}\natexlab{}.
\newblock \showarticletitle{xJsnark: {A} Framework for Efficient Verifiable
  Computation}. In \bibinfo{booktitle}{\emph{{IEEE} S\&P}}.
  \bibinfo{publisher}{{IEEE}}, \bibinfo{pages}{944--961}.
\newblock


\bibitem[\protect\citeauthoryear{Mikkelsen, Damg{\aa}rd, Guldager, Jensen,
  Garcia, Nielsen, Paillier, Pellegrino, Stausholm, Suri, and Zhang}{Mikkelsen
  et~al\mbox{.}}{2015}]%
        {DBLP:books/daglib/p/MikkelsenDGJGNPPSSZ15}
\bibfield{author}{\bibinfo{person}{Gert~L{\ae}ss{\o}e Mikkelsen},
  \bibinfo{person}{Kasper Damg{\aa}rd}, \bibinfo{person}{Hans Guldager},
  \bibinfo{person}{Jonas~Lindstr{\o}m Jensen}, \bibinfo{person}{Jesus~Luna
  Garcia}, \bibinfo{person}{Janus~Dam Nielsen}, \bibinfo{person}{Pascal
  Paillier}, \bibinfo{person}{Giancarlo Pellegrino},
  \bibinfo{person}{Michael~Bladt Stausholm}, \bibinfo{person}{Neeraj Suri},
  {and} \bibinfo{person}{Heng Zhang}.} \bibinfo{year}{2015}\natexlab{}.
\newblock \showarticletitle{Technical Implementation and Feasibility}.
\newblock In \bibinfo{booktitle}{\emph{Attribute-based Credentials for Trust}}.
  \bibinfo{publisher}{Springer}, \bibinfo{pages}{255--317}.
\newblock


\bibitem[\protect\citeauthoryear{M{\"{o}}dersheim, Schlichtkrull, Wagner, More,
  and Alber}{M{\"{o}}dersheim et~al\mbox{.}}{2019}]%
        {DBLP:conf/ifiptm/ModersheimSWMA19}
\bibfield{author}{\bibinfo{person}{Sebastian M{\"{o}}dersheim},
  \bibinfo{person}{Anders Schlichtkrull}, \bibinfo{person}{Georg Wagner},
  \bibinfo{person}{Stefan More}, {and} \bibinfo{person}{Lukas Alber}.}
  \bibinfo{year}{2019}\natexlab{}.
\newblock \showarticletitle{{TPL:} {A} Trust Policy Language}. In
  \bibinfo{booktitle}{\emph{{IFIPTM}}} \emph{(\bibinfo{series}{{IFIP} Advances
  in Information and Communication Technology}, Vol.~\bibinfo{volume}{563})}.
  \bibinfo{publisher}{Springer}, \bibinfo{pages}{209--223}.
\newblock


\bibitem[\protect\citeauthoryear{M{\"{o}}dersheim and Ni}{M{\"{o}}dersheim and
  Ni}{2019}]%
        {DBLP:conf/openidentity/ModersheimN19}
\bibfield{author}{\bibinfo{person}{Sebastian~Alexander M{\"{o}}dersheim} {and}
  \bibinfo{person}{Bihang Ni}.} \bibinfo{year}{2019}\natexlab{}.
\newblock \showarticletitle{{GTPL:} {A} Graphical Trust Policy Language}. In
  \bibinfo{booktitle}{\emph{Open Identity Summit}}
  \emph{(\bibinfo{series}{{LNI}}, Vol.~\bibinfo{volume}{{P-293}})}.
  \bibinfo{publisher}{{GI}}, \bibinfo{pages}{107--118}.
\newblock


\bibitem[\protect\citeauthoryear{More and Alber}{More and Alber}{2022}]%
        {DBLP:conf/IEEEares/MoreA22}
\bibfield{author}{\bibinfo{person}{Stefan More} {and} \bibinfo{person}{Lukas
  Alber}.} \bibinfo{year}{2022}\natexlab{}.
\newblock \showarticletitle{{YOU} {SHALL} {NOT} {COMPUTE} on my Data: Access
  Policies for Privacy-Preserving Data Marketplaces and an Implementation for a
  Distributed Market using {MPC}}. In \bibinfo{booktitle}{\emph{{ARES}}}.
  \bibinfo{publisher}{{ACM}}, \bibinfo{pages}{137:1--137:8}.
\newblock


\bibitem[\protect\citeauthoryear{M{\"{u}}hle, Gr{\"{u}}ner, Gayvoronskaya, and
  Meinel}{M{\"{u}}hle et~al\mbox{.}}{2018}]%
        {DBLP:journals/csr/MuhleGGM18}
\bibfield{author}{\bibinfo{person}{Alexander M{\"{u}}hle},
  \bibinfo{person}{Andreas Gr{\"{u}}ner}, \bibinfo{person}{Tatiana
  Gayvoronskaya}, {and} \bibinfo{person}{Christoph Meinel}.}
  \bibinfo{year}{2018}\natexlab{}.
\newblock \showarticletitle{A survey on essential components of a
  self-sovereign identity}.
\newblock \bibinfo{journal}{\emph{Comput. Sci. Rev.}}  \bibinfo{volume}{30}
  (\bibinfo{year}{2018}), \bibinfo{pages}{80--86}.
\newblock


\bibitem[\protect\citeauthoryear{Podgorelec, Alber, and Zefferer}{Podgorelec
  et~al\mbox{.}}{2022}]%
        {BlazLukasThomasIdentityWalletPaper}
\bibfield{author}{\bibinfo{person}{Blaz Podgorelec}, \bibinfo{person}{Lukas
  Alber}, {and} \bibinfo{person}{Thomas Zefferer}.}
  \bibinfo{year}{2022}\natexlab{}.
\newblock \showarticletitle{What is a (Digital) Identity Wallet? A Systematic
  Literature Review}. In \bibinfo{booktitle}{\emph{The 46th IEEE Computer
  Society Signature Conference on Computers, Software, and Applications
  (COMPSAC 2022)}}.
\newblock


\bibitem[\protect\citeauthoryear{Pr{\"{u}}nster, Ziegler, and
  Palfinger}{Pr{\"{u}}nster et~al\mbox{.}}{2020}]%
        {DBLP:conf/nss/PrunsterZP20}
\bibfield{author}{\bibinfo{person}{Bernd Pr{\"{u}}nster},
  \bibinfo{person}{Dominik Ziegler}, {and} \bibinfo{person}{Gerald Palfinger}.}
  \bibinfo{year}{2020}\natexlab{}.
\newblock \showarticletitle{Multiply, Divide, and Conquer - Making Fully
  Decentralised Access Control a Reality}. In \bibinfo{booktitle}{\emph{{NSS}}}
  \emph{(\bibinfo{series}{{LNCS}}, Vol.~\bibinfo{volume}{12570})}.
  \bibinfo{publisher}{Springer}, \bibinfo{pages}{311--326}.
\newblock


\bibitem[\protect\citeauthoryear{Ro{\ss}nagel}{Ro{\ss}nagel}{2017}]%
        {DBLP:conf/openidentity/Rossnagel17}
\bibfield{author}{\bibinfo{person}{Heiko Ro{\ss}nagel}.}
  \bibinfo{year}{2017}\natexlab{}.
\newblock \showarticletitle{A Mechanism for Discovery and Verification of Trust
  Scheme Memberships: The Lightest Reference Architecture}. In
  \bibinfo{booktitle}{\emph{Open Identity Summit}}
  \emph{(\bibinfo{series}{{LNI}}, Vol.~\bibinfo{volume}{{P-277}})}.
  \bibinfo{publisher}{Gesellschaft f{\"{u}}r Informatik, Bonn},
  \bibinfo{pages}{81--92}.
\newblock


\bibitem[\protect\citeauthoryear{Sabouri, Krontiris, and Rannenberg}{Sabouri
  et~al\mbox{.}}{2012}]%
        {DBLP:conf/trustbus/SabouriKR12}
\bibfield{author}{\bibinfo{person}{Ahmad Sabouri}, \bibinfo{person}{Ioannis
  Krontiris}, {and} \bibinfo{person}{Kai Rannenberg}.}
  \bibinfo{year}{2012}\natexlab{}.
\newblock \showarticletitle{Attribute-Based Credentials for Trust (ABC4Trust)}.
  In \bibinfo{booktitle}{\emph{TrustBus}} \emph{(\bibinfo{series}{{LNCS}},
  Vol.~\bibinfo{volume}{7449})}. \bibinfo{publisher}{Springer},
  \bibinfo{pages}{218--219}.
\newblock


\bibitem[\protect\citeauthoryear{Sporny, Noble, Chadwick, et~al\mbox{.}}{Sporny
  et~al\mbox{.}}{2022}]%
        {w3cVC}
\bibfield{author}{\bibinfo{person}{Manu Sporny}, \bibinfo{person}{Dave Noble,
  Grant~Longley}, \bibinfo{person}{David Chadwick}, {et~al\mbox{.}}}
  \bibinfo{year}{2022}\natexlab{}.
\newblock \bibinfo{booktitle}{\emph{{Verifiable Credentials Data Model 1.1}}}.
\newblock \bibinfo{type}{{W3C} Recommendation}. \bibinfo{institution}{W3C}.
\newblock
\urldef\tempurl%
\url{https://www.w3.org/TR/2022/REC-vc-data-model-20220303//}
\showURL{%
\tempurl}


\bibitem[\protect\citeauthoryear{Wagner, Wagner, More, and Hoffmann}{Wagner
  et~al\mbox{.}}{2019}]%
        {DBLP:conf/openidentity/WagnerWMH19}
\bibfield{author}{\bibinfo{person}{Georg Wagner}, \bibinfo{person}{Sven
  Wagner}, \bibinfo{person}{Stefan More}, {and} \bibinfo{person}{Martin
  Hoffmann}.} \bibinfo{year}{2019}\natexlab{}.
\newblock \showarticletitle{DNS-based Trust Scheme Publication and Discovery}.
  In \bibinfo{booktitle}{\emph{Open Identity Summit}}
  \emph{(\bibinfo{series}{{LNI}}, Vol.~\bibinfo{volume}{{P-293}})}.
  \bibinfo{publisher}{{GI}}, \bibinfo{pages}{49--58}.
\newblock


\bibitem[\protect\citeauthoryear{Wagner and Eckhoff}{Wagner and
  Eckhoff}{2018}]%
        {DBLP:journals/csur/WagnerE18}
\bibfield{author}{\bibinfo{person}{Isabel Wagner} {and} \bibinfo{person}{David
  Eckhoff}.} \bibinfo{year}{2018}\natexlab{}.
\newblock \showarticletitle{Technical Privacy Metrics: {A} Systematic Survey}.
\newblock \bibinfo{journal}{\emph{{ACM} Comput. Surv.}} \bibinfo{volume}{51},
  \bibinfo{number}{3} (\bibinfo{year}{2018}), \bibinfo{pages}{57:1--57:38}.
\newblock


\bibitem[\protect\citeauthoryear{Weinhardt and Omolola}{Weinhardt and
  Omolola}{2019}]%
        {DBLP:conf/icissp/WeinhardtO19}
\bibfield{author}{\bibinfo{person}{Stephanie Weinhardt} {and}
  \bibinfo{person}{Olamide Omolola}.} \bibinfo{year}{2019}\natexlab{}.
\newblock \showarticletitle{Usability of Policy Authoring Tools: {A} Layered
  Approach}. In \bibinfo{booktitle}{\emph{{ICISSP}}}.
  \bibinfo{publisher}{SciTePress}, \bibinfo{pages}{301--308}.
\newblock


\bibitem[\protect\citeauthoryear{Weinhardt and Pierre}{Weinhardt and
  Pierre}{2019}]%
        {DBLP:conf/openidentity/WeinhardtP19}
\bibfield{author}{\bibinfo{person}{Stephanie Weinhardt} {and}
  \bibinfo{person}{Doreen~St. Pierre}.} \bibinfo{year}{2019}\natexlab{}.
\newblock \showarticletitle{Lessons learned - Conducting a User Experience
  evaluation of a Trust Policy Authoring Tool}. In
  \bibinfo{booktitle}{\emph{Open Identity Summit}}
  \emph{(\bibinfo{series}{{LNI}}, Vol.~\bibinfo{volume}{{P-293}})}.
  \bibinfo{publisher}{{GI}}, \bibinfo{pages}{185--190}.
\newblock


\bibitem[\protect\citeauthoryear{Yao}{Yao}{2003}]%
        {DBLP:conf/itrust/Yao03}
\bibfield{author}{\bibinfo{person}{Walt Yao}.} \bibinfo{year}{2003}\natexlab{}.
\newblock \showarticletitle{Fidelis: {A} Policy-Driven Trust Management
  Framework}. In \bibinfo{booktitle}{\emph{iTrust}}
  \emph{(\bibinfo{series}{{LNCS}}, Vol.~\bibinfo{volume}{2692})}.
  \bibinfo{publisher}{Springer}, \bibinfo{pages}{301--317}.
\newblock


\bibitem[\protect\citeauthoryear{Yonezawa, Kobayashi, and Saito}{Yonezawa
  et~al\mbox{.}}{2019}]%
        {draft-yonezawa-pairing-friendly-curves-02}
\bibfield{author}{\bibinfo{person}{Shoko Yonezawa}, \bibinfo{person}{Tetsutaro
  Kobayashi}, {and} \bibinfo{person}{Tsunekazu Saito}.}
  \bibinfo{year}{2019}\natexlab{}.
\newblock \bibinfo{booktitle}{\emph{Pairing-Friendly Curves}}.
\newblock \bibinfo{type}{Internet-Draft}
  draft-yonezawa-pairing-friendly-curves-02. \bibinfo{institution}{IETF
  Secretariat}.
\newblock
\urldef\tempurl%
\url{http://www.ietf.org/internet-drafts/draft-yonezawa-pairing-friendly-curves-02.txt}
\showURL{%
\tempurl}


\bibitem[\protect\citeauthoryear{Yuan and Tong}{Yuan and Tong}{2005}]%
        {DBLP:conf/icws/YuanT05}
\bibfield{author}{\bibinfo{person}{Eric Yuan} {and} \bibinfo{person}{Jin
  Tong}.} \bibinfo{year}{2005}\natexlab{}.
\newblock \showarticletitle{Attributed Based Access Control {(ABAC)} for Web
  Services}. In \bibinfo{booktitle}{\emph{{ICWS}}}.
  \bibinfo{publisher}{{IEEE}}, \bibinfo{pages}{561--569}.
\newblock


\end{thebibliography}

\end{document}